\documentclass[trackchanges]{aastex701}
\usepackage{float}
\usepackage[inkscapeformat=png]{svg}
\usepackage{xspace}
\usepackage[normalem]{ulem}
\usepackage{amsmath}
\usepackage[table]{xcolor}
\usepackage{colortbl} 

\begin{document}

\title{Improving Solar Flare Nowcasting with the Hot Onset Precursor Event (HOPE) Technique}

\author[orcid=0000-0001-5047-5133,sname='Telikicherla']{Anant Telikicherla}
\altaffiliation{Ann and H.J. Smead Department of Aerospace Engineering Sciences, University of Colorado at Boulder}
\affiliation{Laboratory for Atmospheric and Space Physics, University of Colorado at Boulder, 3665 Discovery Dr., Boulder, CO 80303}
\email[show]{anant.telikicherla@lasp.colorado.edu}  

\author[orcid=0000-0002-2308-6797, sname='Woods']{Thomas N. Woods} 
\affiliation{Laboratory for Atmospheric and Space Physics, University of Colorado at Boulder, 3665 Discovery Dr., Boulder, CO 80303}
\email{tom.woods@lasp.colorado.edu}

\author[orcid=0000-0002-1426-6913, sname='Schwab']{Bennet D. Schwab}
\affiliation{Space Sciences Laboratory, University of California Berkeley, 7 Gauss Way, Berkeley, CA 94720}
\email{bennetschwab@berkeley.edu}

\begin{abstract}
This study investigates the statistical behavior of plasma properties during Hot Onset Precursor Events (HOPEs) of solar flares and evaluates their potential for improving flare nowcasting. Two datasets are analyzed: (a) new Soft X-Ray (SXR) spectra of 25 flares (C2.6 to M1.0) obtained from the Dual-zone Aperture X-ray Solar Spectrometer (DAXSS), and (b) SXR irradiance data from 137 flares (C5.0 to X7.1) recorded by the X-Ray Sensor on the Geostationary Operational Environmental Satellite (GOES-XRS). Plasma temperature, emission measure (EM), and low First Ionization Potential (e.g., Mg, Si, Fe) elemental abundance factors (AFs) are derived from DAXSS using Astrophysical Plasma Emission Code model fitting. Isothermal plasma temperature and emission measure are derived from GOES-XRS using the XRS-A/XRS-B ratio method. Results indicate that the HOPE phase exhibits elevated temperatures (10–15 MK) and an order-of-magnitude increase in EM before the impulsive phase. Elemental AFs show a transition from coronal to photospheric values as the flare progresses. Using GOES-XRS data, we develop an improved nowcasting algorithm that detects flares utilizing HOPE signatures. The algorithm is tested across three categories of flares (C5.0–M1.0, M1.0–X1.0, and X1.0+), consistently predicting flare alerts 5–15 minutes ahead of the flare peak. We also explore the possibility of approximate flare magnitude prediction, by calculating correlation between onset parameters and flare peak magnitude. This HOPE-based system shows potential for earlier warnings than current NOAA R3 alerts, which could be useful for High-frequency communication systems operators and targeted flare observation campaigns. 
\end{abstract}

\section{Introduction} \label{sec:intro}
Solar flares generate bursts of electromagnetic radiation in a broad range of wavelengths from gamma rays to radio waves. This radiation rapidly reaches Earth in approximately 8 minutes, where it affects the neutral atmosphere and ionosphere of the Earth, which in turn can disrupt communication and navigation systems \citep{curto_geomagnetic_2020}. Additionally, large eruptive flares are often associated with subsequent Coronal Mass
Ejections (CMEs) and Solar Energetic Particle (SEP) events \citep{kawabata_statistical_2018} that can eject high-energy protons and electrons towards Earth and pose a considerable risk to both satellite systems and astronauts \citep{chancellor_space_2014}. As humanity becomes increasingly reliant on space-based infrastructure and also plans to travel to the Moon (through the upcoming Artemis missions) and Mars, it is crucial to better predict solar flares and device relevant protection strategies. Thus, solar flare prediction on a timescale of a few hours before (i.e. forecasting) and a few minutes before (i.e. nowcasting) are both important for a variety of applications and form a fundamental space weather prediction challenge. In particular, flare nowcasting can be useful for space weather operations such as providing a warning to the aviation industry that they may lose High Frequency (HF) communications in the next few minutes, as well as warnings to astronauts to take shelter. Additionally, nowcasting alerts can also be useful for scientific solar flare observation campaigns, e.g., sounding rockets campaigns that aim to capture the flare’s impulsive phase. 

In this study we focus on solar flare nowcasting using the Hot Onset Precursor Event (HOPE) phenomenon which has been the focus of many recent studies \citep{hudson_hot_2021, battaglia_existence_2023, dasilva_statistical_2023, telikicherla_investigating_2024, hudson_anticipating_2025}. The HOPE phase was originally defined as the “pre-flare interval during which elevated GOES Soft X-Ray (SXR) flux is detected, but prior to the detection of any elevated Hard X-ray (HXR) emission (at $>$25 keV for stronger events, and 12--25 keV for weaker ones) by the Reuven Ramaty High Energy Solar Spectroscopic Imager (RHESSI)'' \citep{hudson_hot_2021}. The HOPE phenomenon has been shown to exist through a variety of independent studies. \cite{hudson_hot_2021} analyzed data from the X-Ray Sensor on the Geostationary Operational Environmental Satellite (GOES-XRS) \citep{woods_goes-r_2024} and data from RHESSI \citep{lin_reuven_2002} to show that the HOPE phase is characterized by flaring plasma having hot temperatures (10--15 MK) as well as an order-of-magnitude increase in plasma emission measure, even before the impulsive phase of the flare begins to ramp up. \citep{battaglia_existence_2023} analyzed multiple flares observed by the Spectrometer Telescope for Imaging X-rays (STIX)\citep{krucker_spectrometertelescope_2020} on Solar Orbiter to confirm the existence of the HOPE phenomena. In addition to measurements from GOES-XRS and STIX, \citep{telikicherla_investigating_2024} investigated the HOPE phenomenon using detailed spectroscopic analysis of the SXR spectra measured by the third-generation Miniature X-ray Solar Spectrometer (MinXSS, also known as the Dual-zone Aperture X-ray Solar Spectrometer (DAXSS) \cite{woods_first_2023}) measurements. Additionally, a statistical study by \cite{dasilva_statistical_2023} has shown that the HOPE phenomenon is very common in flares, with 75\% of flares (in a sample size of 745 flares) showing an onset temperature of more than 8.6 MK. \cite{hudson_anticipating_2025} presented the idea of using HOPE plasma signatures for solar flare nowcasting. This work explores a nowcasting algorithm with promising results and was tested over a  3-day duration with 20 flares as a proof-of-concept. In this study, we perform a statistical analysis of this nowcasting algorithm to establish the accuracy of flare nowcasting using the HOPE plasma parameter signatures. In this study we first extend the statistical analysis of solar flare onsets, by using new data observed by the DAXSS instrument. We analyze temperature, emission measure, and elemental abundance to bring out statistical trends. Subsequently, we test the capability of the flare nowcasting parameters in predicting flare magnitude. Overall, this study addresses the following questions: 
\begin{enumerate} 
    \item What is the statistical variation of the flaring plasma parameters (temperature, emission measure, and abundance factors) during the HOPE phase of solar flares? 
    \item Can the HOPE flaring plasma parameter signatures be used to predict flare peak magnitude and peak time?
\end{enumerate}
The paper is organized as follows; the methodology of DAXSS spectral fitting is described in section \ref{subsec:DAXSS_Methods}. GOES-XRS isothermal temperature and emission measure modelling methodology adopted for this analysis is described in section \ref{subsec:GOES_Methods}. This is followed by discussion of the HOPE plasma signatures (section \ref{subsec:HOPE_sigs}) and the HOPE-based nowcasting algorithm (section \ref{subsec:HOPE_nowcasting}). A statistical analysis of 25 flares observed by DAXSS is presented in section \ref{subsec:DAXSS_analysis}, and 137 flares observed by GOES-XRS is presented in section \ref{subsec:GOES_analysis}. The paper concludes by summarizing the correlation of the flare peak flux magnitude with the onset parameters (section \ref{sec:Results_Conc}) and suggested areas for future work (section \ref{sec:Future_Work}).

\section{Data and Methods} \label{sec:models}
For analyzing the HOPE plasma parameters, two datasets are used in this study which are the DAXSS solar SXR spectra and the GOES-XRS irradiance. This section describes data products as well as the model fitting methodology employed for obtaining the plasma parameters from the instrument measurements. This is followed by a discussion of the typical HOPE plasma parameter signatures for an example solar flare. Lastly, the flare nowcasting algorithm adopted for this study is discussed in detail. 

\subsection{DAXSS SXR Spectrum} \label{subsec:DAXSS_Methods}
\begin{figure*}[ht!]
\centering
\includegraphics[width = \textwidth]{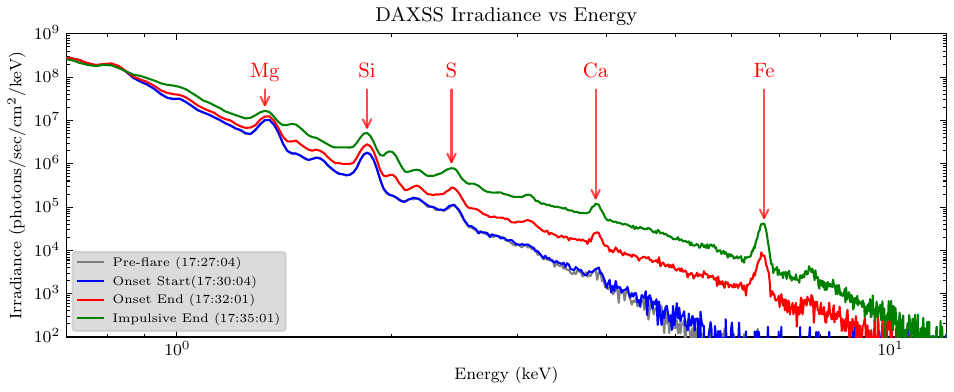}
\caption{DAXSS Spectra during times of a M1.0 flare on 2022-08-15. The pre-flare spectra is shown in grey, onset start in blue, onset end in red, and impulsive end in green. The line-complexes of 5 low-FIP elements (Mg, Si, S, Ca, and Fe) are also marked with red arrows. Note: Sulphur has a FIP of 10.36 eV and is at the border between low-FIP and high-FIP elements. The Mg feature has significant line-blending with Fe and Al above 10 MK, additional details of various spectral features in the DAXSS SXR range are described in appendix A of \cite{woods_first_2023}}.
\label{fig:spectra_variation}
\end{figure*}

The DAXSS instrument provides SXR spectra in the energy range of 0.7 to 12 keV, with an energy resolution of 0.05 keV at 1 keV, and a cadence of 9 seconds \citep{woods_first_2023}. The instrument is onboard the INSPIRESat-1 \citep{chandran_inspiresat-1_2021} spacecraft and has down-linked data flare data from February 2022 to December 2023. To determine flaring plasma parameters, spectral fitting is performed using PyXSPEC, which is the Python interface to the XSPEC \citep{arnaud_xspec_1996} spectral-fitting program. The spectral fitting procedure is described in detail in \cite{telikicherla_investigating_2024} and key aspects are summarized here. The measured spectra are fit to the Astrophysical Plasma Emission Code (APEC) vvapec model based on the AtomDB Atomic Database \citep{foster_updated_2012}, to compute the plasma temperature, emission measure, and elemental abundance factor ratios. For this study, based on the model fitting and comparisons explained in previous studies by \citep{telikicherla_investigating_2024, woods_first_2023, nagasawa_study_2022}, we use a 3-temperature model in which the elemental abundances corresponding to the three temperature components are tied together. For each spectra, the energy upper limit for model fitting is chosen according to the counts of the spectra i.e. the energy limit is where the counts are less than 2 per integration time (which is 9 seconds). Based on this upper energy limit, different elemental abundance factors are included in the model. These include low first-ionization potential (FIP) elements like Mg, Si, and Fe. Further details of the various line complexes in the DAXSS spectra are described in \cite{woods_first_2023}. As an example, the SXR spectra for pre-flare, start and end of onset, and end of impulsive phase for a M1.0 flare on 2022-08-15 is shown in Figure \ref{fig:spectra_variation}. The y-axis is in irradiance units (photons/sec/cm$^2$/keV) which considers the responsivity of the instrument to convert from count rates to irradiance. As depicted in the figure, as the flare progresses from pre-flare to onset, the slope of the continuum changes, indicating an increase in plasma temperature. Additionally, hotter Ca and Fe line-complexes become more prominent as the flare progresses from onset to impulsive phase. Detailed temporal evolution of plasma parameters obtained from these spectra are described in section \ref{subsec:HOPE_sigs}.

\subsection{GOES XRS Irradiance} \label{subsec:GOES_Methods}
The XRS instrument onboard GOES provides full‐disk solar SXR irradiance in two broadband wavelength intervals of (a) 0.05–0.4 nm, called XRS‐A; and (b) 0.1–0.8 nm, called XRS‐B \citep{woods_goes-r_2024}. These are available at a time cadence of 1 second. Coronal plasma temperature can be determined using a combination of XRS‐A and B data. In this study, we adopt the widely used technique of computing iso-thermal plasma temperature and emission measure by computing the ratio of the XRS‐A irradiance to the XRS‐B (denoted by R = XRS-A/XRS-B). Based on this ratio, the plasma temperature and emission measure can be computed using a spectral model for the solar irradiance \citep{white_updated_2005, woods_first_2023}, as follows:
\begin{equation}
\begin{split}
T_{XRS}(MK) &= 2.7460 + 129.47\cdot R - 966.28 \cdot R^2 + 5517.5 \cdot R^3 - 1.8664*10^4 \cdot R^4\\
&\quad + 3.5951*10^4 \cdot R^5 - 3.6099*10^4 \cdot R^6 + 1.4687*10^4 \cdot R^7
\end{split}
\end{equation}
\begin{equation}
\begin{split}
XRS\_B_{model} &= 6.9469 - 6.0827 \cdot T_{XRS} + 1.7364 \cdot T_{XRS}^2 - 0.15594 \cdot T_{XRS}^3\\
&\quad + 6.7848*10^{-3}\cdot T_{XRS}^4 - 1.4446*10^{-4} \cdot T_{XRS}^5 + 1.2089*10^{-4} \cdot T_{XRS}^6
\end{split}
\end{equation}

\begin{equation}
    EM_{XRS}(10^{49}cm^{-3}) = \frac{XRS\_B_{measure}}{XRS\_B_{model}} * 10^{-6}
\end{equation}
These equations are for GOES-16 data where the XRS-A and B channel measurements are in Wm$^{-2}$ units. Additionally, we note that these equations are not available in the SolarSoft library, and thus the temperature and emission measure results are expected to have slight variations. Based on the methods described here to determine plasma parameters, the next section gives an example of the HOPE plasma parameter signatures computed using data from both the DAXSS and GOES-XRS instruments.

\subsection{HOPE Plasma Parameter Signatures} \label{subsec:HOPE_sigs}
\begin{figure*}[hb!]
\centering
\includegraphics[width = \textwidth]{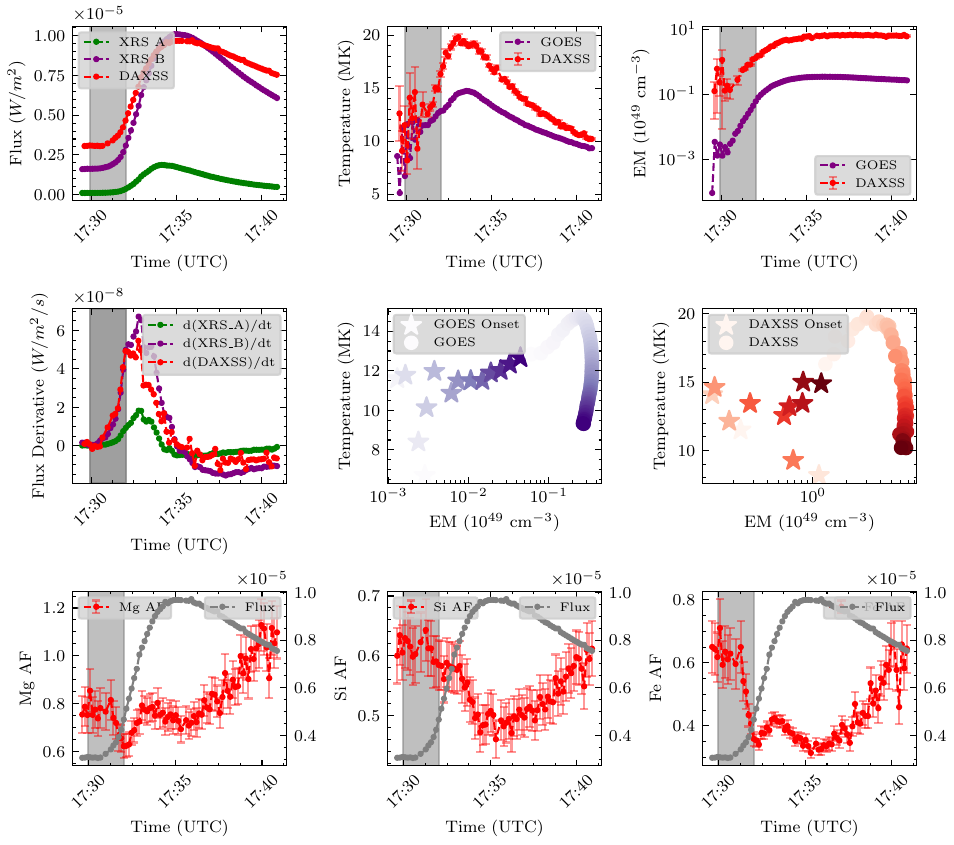}
\caption{Plasma parameter signatures during the HOPE phase of a M1.0 flare on 2022-08-15. The upper left panel shows the SXR flux measured by XRS-A, XRS-B, and DAXSS. Upper middle and upper right panel show the plasma temperatures and emission mission measure respectively. The middle-left row shows the flux derivative, of XRS-A, XRS-B, and DAXSS. The middle and right panel of the middle row show the characteristic temperature vs emission measure loop, from GOES and DAXSS respectively. Lastly, the bottom panels show the Mg, Si, and Fe abundance factor variations respectively.}
\label{fig:HOPE_signatures}
\end{figure*}

The temporal variations of plasma temperature, emission measure, and elemental abundances factors during the HOPE phase of an example M1.0 flare on 2022-08-15 are depicted in Figure \ref{fig:HOPE_signatures}. The grey shaded region in the panels denotes the time corresponding to the HOPE phase of the flare, prior to the impulsive phase. 
The upper left panel shows the flux measured by the GOES XRS Channels A and B, as well as the DAXSS instrument (note the DAXSS measurements here are total counts per second, which are scaled to appear in the same plot as flux). The middle left panel shows the flux derivatives of XRS-A, XRS-B, and DAXSS. These are calculated by computing the time-derivative at each point of the XRS-A, XRS-B, and DAXSS timeseries. The upper middle panel shows the plasma temperature as obtained by both DAXSS and GOES-XRS, showing that the plasma temperature increases significantly (to a range of 8-15 MK) during the HOPE phase. As described in section \ref{subsec:DAXSS_Methods}, a three temperature model is used for DAXSS spectral fitting, and the temperature and emission measure corresponding to the hottest temperature component are plotted here. The uncertainties in plasma parameters derived from DAXSS measurements (depicted with red error bars in the Temperature, emission measure, and abundance factor plots) are derived using the one-sigma confidence interval obtained from PyXSPEC. To determine the GOES temperature, first a pre-flare background level is subtracted and then the residual flux ratio is used to compute isothermal plasma temperature and emission measure as described in section \ref{subsec:GOES_Methods}. A simple method for estimating uncertainties for the GOES-XRS derived plasma parameters, is by assuming a 5\% variation in the XRS-A/XRS-B ratio, and estimating the uncertainty in the isothermal temperature and emission measure. Using this method, considering a temperature range of 5-30 MK, the fractional uncertainties in temperature and emission measure range from approximately 1.94\% to 7.12\% and 1.56\% to 5.84\% respectively. For example, a 5\% variation in the flux ratio of 0.01 would lead to a temperature of $10.00 \pm 0.21 MK$ and an emission measure of $5.42\cdot10^{-2} \pm 1.59\cdot10^{-3}$ (in units of $10^{49}$ $cm^{-3}$). The upper right panel shows the plasma emission measure and indicates an order-of-magnitude increase during the HOPE phase. This is also depicted in the middle row center and right plots, that show the characteristic plasma temperature vs emission measure loop for GOES and DAXSS respectively. In the temperature vs emission measure loop plots, the onset phase is marked using stars and the main flare using circle symbols. The onset phase starts when the plasma temperatures increases to the 10-15 MK range, while the SXR flux is close to the background level. The onset phase ends as the SXR flux increases rapidly during the impulsive phase, indicated by an increase in the flux derivative. Additionally,  a gradient from light to dark is applied both to the onset phase and impulsive phase symbols, to indicated temporal variation (light to dark represents advancing time). The bottom row shows the abundance factors (AFs) of Mg, Si, and Fe respectively. An AF value of 1 indicates Feldman Standard Extended Coronal (FSEC) abundances. A reduction from 1 towards 0.25, indicates a value changing from coronal towards photospheric values. Mg, Si and Fe AFs all show the trend of falling below the pre-flare values during the HOPE phase. Mg and Fe show some receovery after the HOPE phase, which is then followed by decrease and recovery in the impulsive and gradual phases, respectively. 
\subsection{HOPE-based Flare Nowcasting Algorithm} \label{subsec:HOPE_nowcasting}
This section describes the algorithm for flare nowcasting, based on the plasma parameter signatures observed during the HOPE-phase.  This algorithm is based on the nowcasting technique described by \cite{hudson_anticipating_2025}. In this method, first a running-difference of the measured flux in both XRS channels is calculated (denoted by $\Delta XRSA, \Delta XRSB$) based on a particular time difference value (denoted by $t_{diff}$). The plasma temperature and emission measure is then computed from this running-difference flux using the ratio method described in section \ref{subsec:GOES_Methods}. Then the following conditions are checked:

\begin{equation}
    \Delta EM > EM_{Threshold}
\end{equation}
\begin{equation}
    \Delta T(MK) > \Delta T_{min} 
\end{equation}
Once these conditions are satisfied a flare trigger is generated. The various tunable parameters of this nowcasting algorithm are summarized in Table \ref{tab:goes_fai_parameters}. In addition to the emission measure threshold and the minimum temperature, the integration time of the XRS flux measurements, and the time value used for running difference are also tunable parameters for this algorithm. Appendix \ref{sec:appendix} describes the parameter selection methodology followed for choosing the algorithm parameter values for best performance, which are listed in the selected value column of Table \ref{tab:goes_fai_parameters}.
\begin{deluxetable*}{ccccc}[hb!]
\tablewidth{0pt}
\tablecaption{HOPE-based flare nowcasting algorithm parameters \label{tab:goes_fai_parameters}}
\tablehead{
\colhead{S. No.} & \colhead{Parameter} & \colhead{Symbol} & \colhead{Selected Value} & \colhead{Unit}
}
\startdata
1 & XRS Flux Integration Time   & $t_{\mathrm{int}}$        & 60                   & seconds \\
2 & Running-difference Time     & $t_{\mathrm{diff}}$       & 180                  & seconds \\
3 & Emission Measure Threshold  & $\Delta EM$               & $5 \times 10^{-3}$   & $10^{49}\,\mathrm{cm}^{-3}$ \\
4 & Temperature Threshold       & $\Delta T_{\mathrm{min}}$ & 5                    & MK \\
\enddata
\end{deluxetable*} 

\begin{figure*}[hb!]
\centering
\includegraphics[width = \textwidth]{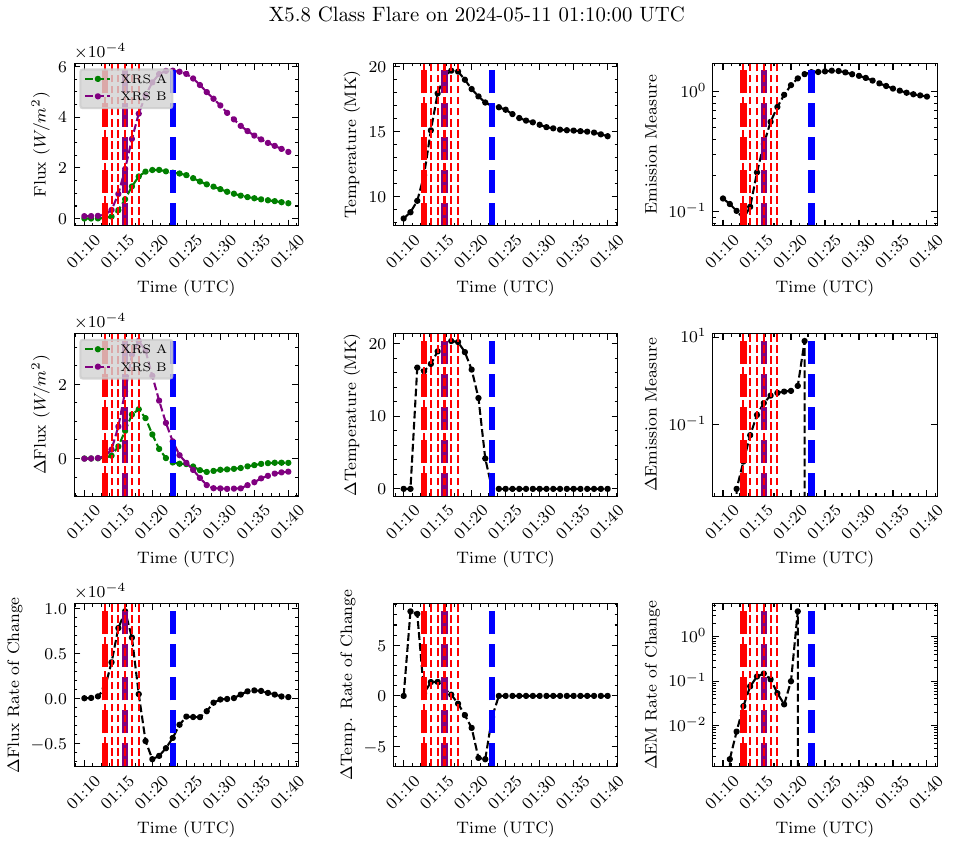}
\caption{Example of the onset detection algorithm using GOES-XRS data, as applied to a X5.8 flare on 2024-05-11 during the May 2024 Gannon storm. The first column shows plots of XRS-A and XRS-B flux, running-difference flux (using 3 minute intervals), and rate of change of running-difference flux in the upper, middle, and bottom rows respectively. The second column depicts the isothermal plasma temperature in the first row, temperature determined from running difference flux in the second row, and rate of change of running-difference temperatures in the third row. Lastly, the third column shows, isothermal plasma emission measure, emission measure determined from running difference flux, and the rate of change of this running-difference emission measure in the first, second, and third row respectively. The thin red vertical lines indicate the triggers obtained from HOPE nowcasting algorithm, and the thick vertical red line indicates the first trigger. The dashed vertical purple lines indicate the local maximum for the $\Delta EM$ rate of change, and the vertical blue lines indicate the flare peak time. }
\label{fig:trigger_algorithm}
\end{figure*}

Figure \ref{fig:trigger_algorithm} shows the flare nowcasting algorithm applied using data from GOES-XRS to an example X5.8 flare, that was one of the flares during the May 2024 Gannon Storm. The XRS-A and XRS-B fluxes are shown in the upper left panel. The middle and right column of the first row show the isothermal temperature and emission measure as determined from the XRS-A and XRS-B fluxes, using the ratio method but without background subtraction.   
The second row shows the running difference fluxes ($\Delta$XRS-A and $\Delta$XRS-B) in the left column, and corresponding temperature ($\Delta T$) and emission measure ($\Delta EM$) in the center and right columns respectively. The temperature and emission measures obtained from the running difference fluxes are set to 0 in case either of the running differences are negative, which happens after the impulsive phase of the flare, where $\Delta$XRS-A and/or $\Delta$XRS-B are negative. In this example, the algorithm is triggered when  $\Delta EM>5\cdot10^{-3}$ (in units of $10^{49}\,\mathrm{cm}^{-3}$) and $\Delta T > 5$ MK. Thin vertical red lines are included for each data point that satisfies this flare trigger condition, and the thick vertical red line indicates the first trigger. The bottom row represents the rate of change of the running-difference parameters, calculated by taking the time-derivative of the running-difference parameters.  The left column shows the rate of change of the running-difference flux, the middle column shows the rate of change of the running-difference temperature, and the right column shows the rate of change of the running-difference emission measure. It was observed for most flares that the rate of change of the running-difference emission measure has a local maximum prior to the flux peak (denoted by the vertical blue lines). This local maximum is denoted by vertical purple lines. Since the running-difference represents a form of the first-derivative of the parameters, the derivative of the running-difference represents the second derivative of each parameter. Therefore, a peak in the second derivative could represent the time when flare heating has reached its maximum heating rate. The value of $\Delta EM$ at this local maximum is considered as a parameter for flare magnitude prediction (denoted by $\Delta EM$ at $\Delta^2 EM$ local maximum), and is discussed further in section \ref{subsec:GOES_analysis}. Based on this algorithm, statistical studies were performed to analyze the properties of flare onsets as well as to quantify the accuracy of generating solar flare alerts and predicting flare magnitude, which are described in the subsequent sections. Because DAXSS data saturates at a M1 level, it is primarily used for spectroscopy and determination of trends in HOPE plasma properties. GOES-XRS data on the other hand has a much broader flux range and is used in this study for analysis of the flare nowcasting algorithm. 

\section{Observation and Discussion}\label{sec:Obsv_Disc}
This section is divided into two subsections, the first describes the statistical analysis performed on DAXSS flare data and the second describes the statistical analysis of flare nowcasting performed on GOES flare data. The DAXSS data analysis consists of 25 flares observed during the time frames of March to October 2022. These flares range from C2.6 to M1.0 class, as the DAXSS instrument saturates above this level. Because DAXSS observations are limited to smaller flares, the main purpose of analysis is to understand the plasma parameter trends (temperature, emission measure, and abundance factors). The main objective of the second subsection is to understand the onset plasma parameter statistics and test the flare nowcasting algorithm on a dataset consisting of a broad range of flare magnitudes. For this study, we consider 137 flares ranging from a flare class of C5.0 to X7.1. Flares below the C5.0 class are not considered, as their impact on Earth's neutral atmosphere and ionosphere is considerably less, and thus are not very critical for space weather operations.


\subsection{Analysis of DAXSS Flare Data}\label{subsec:DAXSS_analysis}
\begin{figure*}[ht!]
\centering
\includegraphics[width = \textwidth]{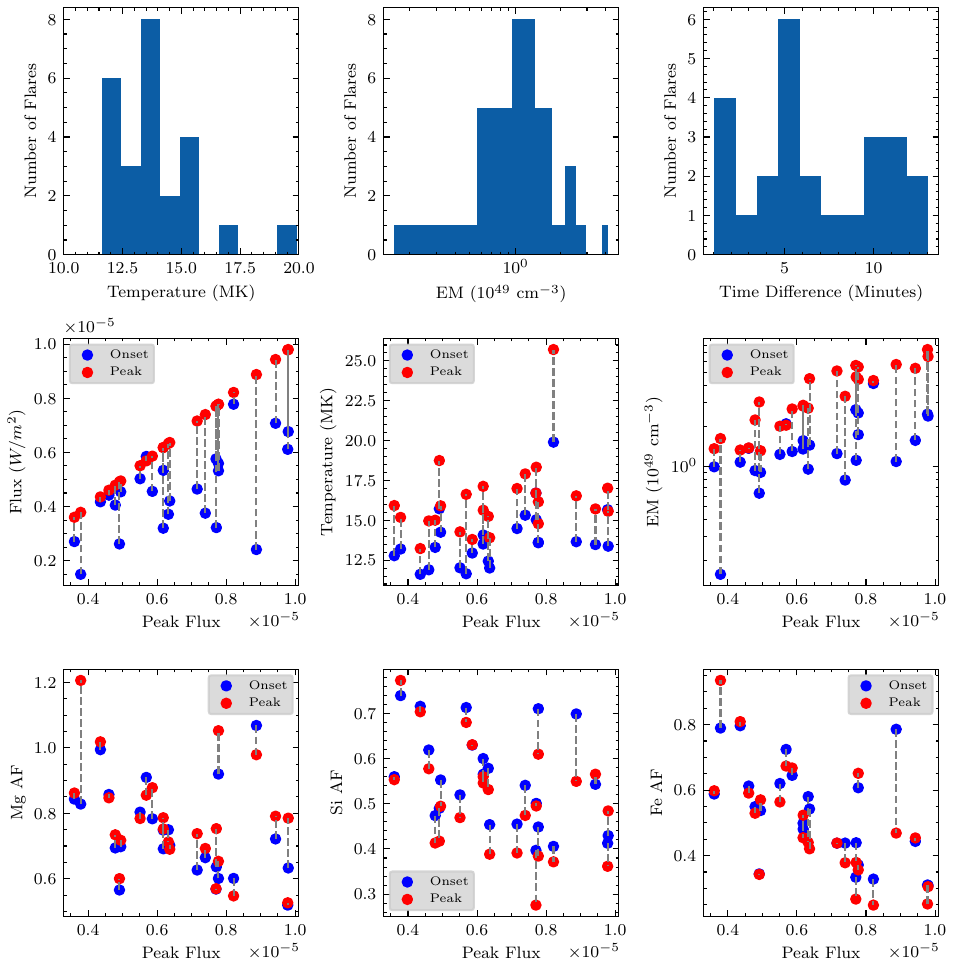}
\caption{Statistical variation of plasma properties during the HOPE phase of 25 flares (ranging from C2.6 to M1.0 class) measured by the DAXSS instrument. Histograms of plasma temperature, emission measure, and time difference between the first flare onset trigger and flare peak are shown in the left, middle, and right plots of the first row. The second row shows the variation of SXR flux, temperature, and emission measure with the peak flux (in $Wm^{-2}$), during the onset (blue) and flare-peak (red). Gray dotted lines are drawn to guide the eye between onset and peak parameters of the same flare. Similarly, the third row depicts the abundance factor variation of Mg, Si, and Fe with the peak flux (in $Wm^{-2}$) in the left, middle, and right panels respectively. An abundance factor of 1 indicates Feldman Standard Extended Coronal (FSEC) abundances.}
\label{fig:daxss_statitics}
\end{figure*}

Figure \ref{fig:daxss_statitics} shows the distribution of plasma temperature, emission measure, and abundance factors obtained from DAXSS model fitting. The first row shows the temperature and emission measure distribution during flare onset in the left and middle panels, respectively. For this sample of 25 flares, the mean plasma temperature during the onset is 13.85 MK with a standard deviation of 1.81 MK. Additionally, the emission measure has a mean value of 1.51, with a standard deviation of 0.81, in $10^{49}$$cm^{-3}$ units. The upper right panel shows the distribution of the time difference between the first flare onset trigger and flare peak, with a mean time of 6.56 minutes and a standard deviation of 3.67. The middle row shows the variation of flux, plasma temperature, and emission measure in the left, middle, and right panels respectively. Note that the x-axis is the DAXSS total counts per second scaled to be approximately the same order of magnitude as the GOES-XRS flux measurements. The onset parameters are plotted using blue symbols and the flare peak parameters are plotted using red symbols. Gray dotted lines are drawn to guide the eye between onset and peak parameters of the same flare.  The plots show a weak correlation between the onset parameters and the flare peak flux magnitude; indicating higher onset flux, onset temperature and emission measures as the flare peak flux (class) increases. A detailed analysis of the correlation of the onset properties to the main flare peak flux are performed using the using the GOES-XRS dataset which provides a wider range of flare measurements for significant estimation of statistical metrics, as DAXSS saturates above the M1 level. The bottom row shows the onset and flare-peak abundance factors of Mg, Si, and Fe in the left, middle, and right panels respectively. In these plots an abundance factor value of 1 indicates Feldman Standard Extended Coronal (FSEC) abundances. A reduction from 1 towards 0.25, indicates a value changing from coronal towards photospheric values. 
These plots indicate a trend of higher abundance factor depletion for flares with larger flare peak flux (i.e., larger flare class). For Si, 21 flares (84\%) have a peak AF lesser than the onset AF, for Fe 16 flares (64\%) have a peak AF lesser than the onset AF, however for Mg only 7 flares (28\%) have a peak AF less than the onset AF. The reason for this difference is beacase of line-blending in the Mg features. Appendix A from \cite{woods_first_2023} gives a detailed description of the various spectral features in the DAXSS SXR range. Mg features are the strongest below 10 MK, but have high contributions from Fe and Al (which is not a low-FIP element) as the temperature increases. On the other hand, Si and Fe features are more ``pure" with lesser line blending as the temperature increases. This suggests that Si and Fe features are better to study the FIP effect than the Mg features. The observation of the AFs falling from the coronal value to photospheric values indicate the importance of chromospheric evaporation as a heating mechanism during the HOPE phase of the flares. Additionally, larger depletion for larger flares indicates that more heated plasma from the chromosphere rises towards the corona for larger flare classes. 

\subsection{Statistical Analysis of HOPE-based nowcasting using GOES-XRS Data}\label{subsec:GOES_analysis}
\begin{figure*}[hb!]
\centering
\includegraphics[width = \textwidth]{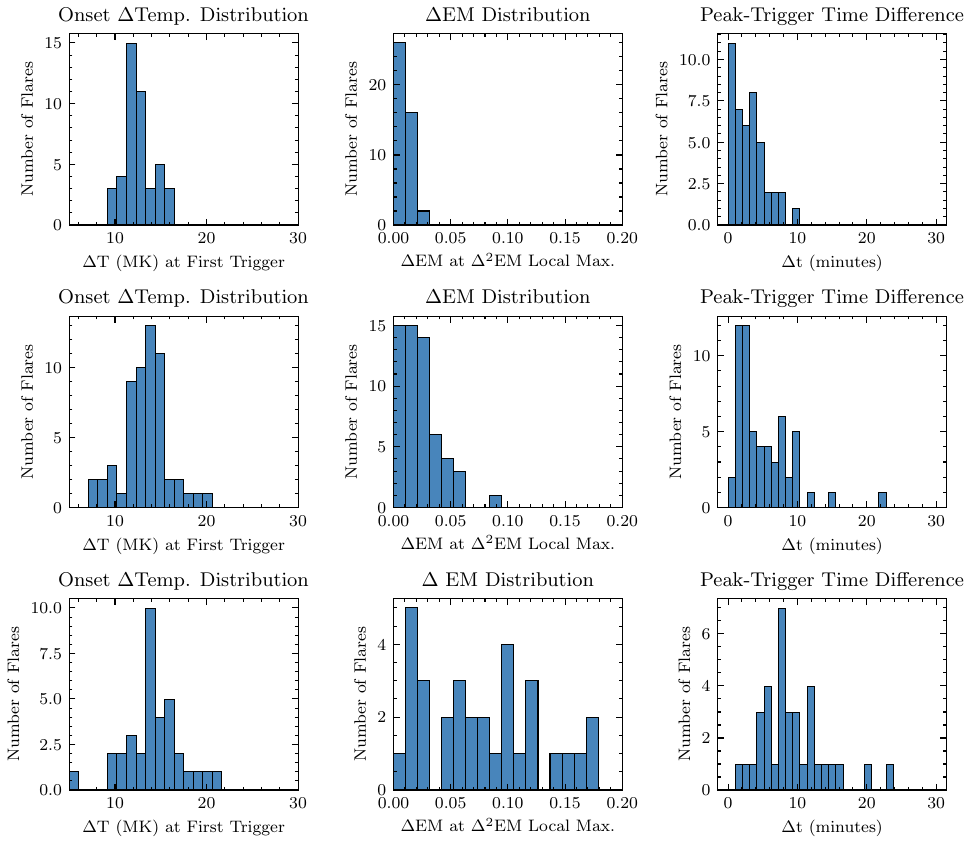}
\caption{Statistical variation of plasma parameters during the HOPE phase of 137 flare measured by the GOES XRS instrument. The first row shows the results for 44 category (a) flares (C5.0 to M1.0). The second row shows results for 58 category (b) flares (M1.0 to X1.0), and the third row shows 35 category (c) flares (X1.0 and above). The first column is a histogram of the running-difference temperature ($\Delta T$) at the first trigger. The second column is a histogram of the running-difference emission measure ($\Delta EM$) at the local maximum of the ($\Delta^2EM$). The third column shows a histogram of the time difference between the first HOPE trigger and the flare peak.}
\label{fig:goes_stats}
\end{figure*}

This section describes the analysis of GOES-XRS data in order to determine the performance of the HOPE-based flare nowcasting algorithm over a wide range of flare classes. Thus, a case study based on a dataset comprising of 137 flares ranging from C5.0 to X7.1 flare class is described below. The flares are divided into three categories according to peak magnitude which are (a) C5.0 to M1.0 (44 flares), (b) M1.0 to X1.0 (58 flares), and (c) greater than X1.0 (35 flares). Fig \ref{fig:goes_stats} shows the results from this case study.
The first row represents category (a) flares, second row represents category (b) flares, and third row represents category (c) flares.  The first column depicts a histogram of the running-difference temperature ($\Delta T$) at the first onset trigger point. The second column depicts the running-difference emission measure ($\Delta EM$) at local maximum of the $\Delta^2 EM$, and the third column depicts the time difference (in minutes) between the first HOPE trigger and the flare peak. Note that the running difference emission measure ($\Delta EM$) is not depicted at the first trigger point, as it is itself a trigger parameter and that would bias the histogram statistics to the trigger threshold. For each category of flares the mean and standard deviation of each parameter are listed in Table \ref{tab:goes_onset_stats}. In this table the standard deviation is the spread of the parameters from the mean value and is different from the uncertainties discussed in section \ref{subsec:HOPE_sigs}. It can be observed that the running-difference temperature ($\Delta T$) has a slight increasing trend with flare class, although most of them are in the 10-15 MK range. The running difference emission measure ($\Delta EM$) at the local maximum of the ($\Delta^2 EM$) also shows an increasing trend with the flare peak magnitude. Lastly the time difference between the onset trigger and the peak flare shows a similar increasing trend with increasing flare class, with the mean value of the time difference being $9.38 \pm 4.49$ minutes for category (c) flares. This indicates that the HOPE alert can come approximately 5-15 minutes before the flare peak for X-class flares.
\begin{deluxetable*}{cccc}[ht!]
\tablewidth{0pt}
\tablecaption{Flare onset parameters for the three categories of flares analyzed. \label{tab:goes_onset_stats}}
\tablehead{
\colhead{Parameter} & 
\colhead{Category (a): C5.0--M1.0} & 
\colhead{Category (b): M1.0--X1.0} & 
\colhead{Category (c): $>$X1.0}
}
\startdata
Temperature $\Delta T$ (MK) & $12.62 \pm 1.62$ & $13.32 \pm 2.52$ & $14.19 \pm 2.92$ \\
Emission Measure $\Delta EM$ ($10^{49}\,\mathrm{cm}^{-3}$) & $1.03 \cdot 10^{-2} \pm 5.30 \cdot 10^{-3} $ & $2.38 \cdot 10^{-2} \pm 1.62 \cdot 10^{-2}$ & $9.41 \cdot 10^{-2} \pm 7.21 \cdot 10^{-2}$ \\
Time Difference $\Delta t$ (minutes) & $3.46 \pm 2.26$ & $5.39 \pm 3.86$ & $9.38 \pm 4.49$ \\
\enddata
\tablecomments{The table lists the  mean and standard deviation for the running-difference plasma temperature ($\Delta T$) at the first trigger, running-difference emission measure ($\Delta EM$) at the local maximum of ($\Delta^2 EM$), and the time difference between the first trigger and the flare peak. The standard deviation is the spread of the parameters from the mean value and is different from the uncertainties discussed in section \ref{subsec:HOPE_sigs}.}
\end{deluxetable*}

\section{Results and Conclusion}\label{sec:Results_Conc}

\begin{figure*}[!htbp]
\centering
\includegraphics[width = \textwidth]{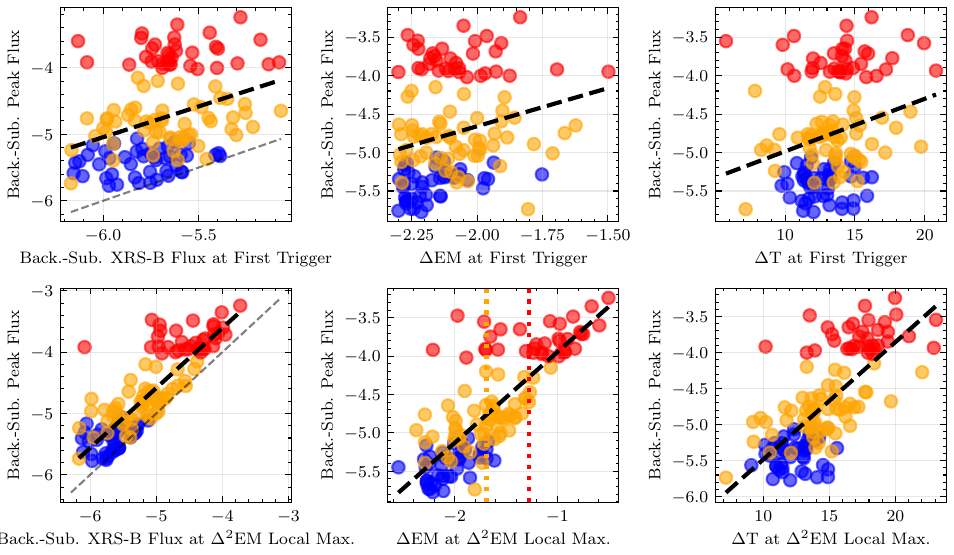}
\caption{Correlation of onset parameters with the logarithm of background subtracted peak flux (in W$m^{-2}$). First row depicts the logarithm of the background subtracted peak flux, logarithm of the running-difference emission measure ($\Delta EM$), and running-difference temperature ($\Delta T$) at the first trigger in the first, second, and third columns respectively. The second row shows the same parameters at the local maximum of the derivative of the running difference emission measure ($\Delta^2 EM$). C class flares are depicted using blue symbols, M class using yellow symbols, and X class flares using red symbols. The dashed gray lines in the first column indicate the y=x line to show how much lower the onset flux is compared to the peak flux. The dashed black lines show the least-square fit lines to the parameters. In the lower middle plot the 95th percentile value of the running difference emission measure for C-class flares is depicted using a yellow vertical dashed line, and 95th percentile of the M-class flares is depicted using the red vertical dashed line.}
\label{fig:mag_predict}
\end{figure*}

In the previous section, the statistical variation of the onset trigger for a wide range of flares (C5.0 to X7.1) were presented. In this section, the possibility of predicting flare peak magnitude is explored.  Fig \ref{fig:mag_predict} shows the correlation of different onset parameters with the background subtracted flare peak flux magnitude, for all 137 flare analyzed in the previous section. 
Category (a) flares are represented by blue symbols, category (b) by yellow symbols, and category (c) by red symbols. The first row shows the correlations at the first trigger. The y-axis in all these plots is the logarithm of the background subtracted peak flux, i.e. an M class flare with a magnitude between $10^{-5}$ W$m^{-2}$ and $10^{-4}$ W$m^{-2}$, would be between -5 and -4 on the plot. As shown in the top-left panel, the background subtracted XRS-B flux is at low level (less than $10^{-5}$ $Wm^{-2}$) at the first trigger, for all three categories of flares. The dashed grey line depicts the y=x line, to guide the eye, and show that the flux at the first trigger is considerably lower than the peak flux for bigger flares (M and X class). The second and third columns of the first row show the correlation of $\Delta EM$ and $\Delta T$ with the flare peak magnitude at the first trigger, respectively. Both these parameters show a weak correlation with flare magnitude at the onset. For all plots, dashed black lines indicate the least-square fit line to the data. The second row shows the correlation of the background subtracted XRS-B flux at the local maximum of the $\Delta^2EM$. This shows a stronger correlation with peak flux, but is also closer to the y=x line, indicating that the flux at the local maximum of $\Delta^2EM$ is close to the flare peak. The middle and right panel in the bottom row show the correlations between the $\Delta EM$ and $\Delta T$ at the local maximum of $\Delta^2EM$. The parameters of the second row show a stronger correlation to the flare peak magnitude, indicating that they can be used for a more accurate flare magnitude prediction, with the caveat that this is much closer to the flare peak in time (i.e. the local maximum of the $\Delta^2EM$ can be approximately 5 minutes after the first trigger, and thus closer to the flare peak). The correlations obtained between the plasma parameters and the flare peak magnitude are summarized in Table \ref{tab:flare_peak_corr_combined}. The table lists the slope, intercept and $R^2$ value of the least-square fit linear equations relating the onset parameters to the flare peak parameters. The $R^2$ value for the first trigger are low (approximately 0.1) indicating a poor correlation at the first trigger. However, the $R^2$ value at the local maximum of $\Delta^2EM$ are higher (0.70 for XRS-B flux) showing that the flare magnitude prediction improves from the first trigger to the trigger at the local maximum of $\Delta^2EM$. 
\begin{deluxetable*}{ccccccc}[!hb]
\tablewidth{0pt} 
\tablecaption{Linear Fit Results of onset parameters \label{tab:flare_peak_corr_combined}}
\tablehead{
\colhead{Parameter} &
\multicolumn{3}{c}{At First Trigger} &
\multicolumn{3}{c}{At $\Delta^2EM$ Local Maximum} \\
\colhead{} &
\colhead{Slope} & \colhead{Intercept} & \colhead{$R^2$} &
\colhead{Slope} & \colhead{Intercept} & \colhead{$R^2$}
}
\startdata
XRS-B        & 0.92 & 0.50  & 0.11 & 0.98 & 0.31  & 0.70 \\
$\Delta EM$  & 0.99 & -2.68 & 0.05 & 1.18 & -2.76 & 0.62 \\
$\Delta T$   & 0.07 & -5.67 & 0.06 & 0.16 & -7.11 & 0.52 \\
\enddata
\tablecomments{The table shows the linear fit results between HOPE parameters (background subtracted XRS-B flux, running-difference temperature ($\Delta T)$, and running-difference emission measure ($\Delta EM$)) with background-subtracted flare peak flux at two times: the First HOPE Trigger and at the $\Delta^2EM$ Local Maximum.}
\end{deluxetable*}

\begin{figure*}[!htbp]
\centering
\includegraphics[width = \textwidth]{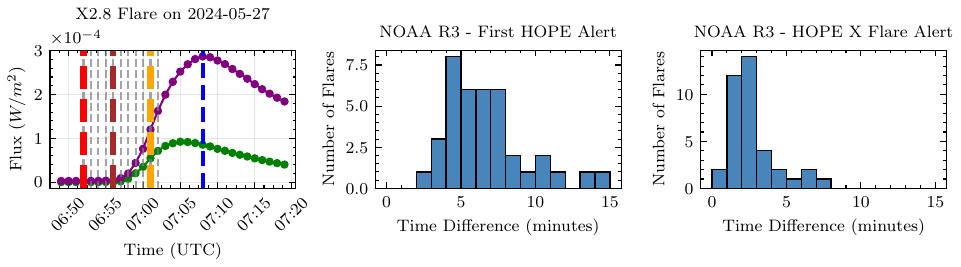}
\caption{Application of the HOPE based nowcasting methodology developed in this study to generate early Radio Blackout alerts. The left panel shows an example timeseries of the GOES XRS-A (green) and XRS-B (purple) flux for a X2.8 flare on 2024-05-27. The vertical red line indicates the first HOPE alert and the vertical grey lines represent the subsequent HOPE alerts. The vertical brown line shows the HOPE X-class flare alert, and the vertical yellow line shows the NOAA R3 Radio Blackout alert. The middle panel shows a histogram of the time difference between the NOAA R3 alert and the first HOPE alert for all category (c) X1.0 and above flares described in the previous section. The right panel shows a histogram of the time difference between the NOAA R3 radio blackout alert and the HOPE X-class flare alert.}
\label{fig:alerts}
\end{figure*}

One goal of this nowcasting technique is to produce earlier Radio-blackout alerts. Currently, NOAA raises 5 levels of alerts based on the flux level monitered by GOES-XRS channel B. These are R1 (minor: triggered at a M1 flux level), R2 (moderate: triggered at a M5 flux level), R3 (strong: triggered at a X1 flux level), R4 (severe: triggered at a X10 flux level), and R5 (extreme: triggered at a X20 flux level). By applying the HOPE nowcasting algorithm such alerts can be generated earlier. Figure \ref{fig:alerts} shows the application of the HOPE nowcasting methodology to generate early R3 radio-blackout alerts. The left panel shows an example timeseries of an X2.8 flare on 2024-05-07 to illustrate the methodology. The XRS-A flux is shown in green and XRS-B flux is shown in purple. The vertical red line shows the first HOPE alert according to the trigger conditions described in this study, and the vertical gray lines show the subsequent alerts. The vertical brown line indicates when a HOPE alert for an X-class flare is raised. This is computed by checking the value of the running-difference emission measure crossing a minimum threshold. The minimum threshold is computed using the correlation plots in Figure \ref{fig:mag_predict}, by taking the 95th percentile level of ($\Delta EM$) for category (b) flares (M1.0 - X1.0). Thus, a value greater than this threshold would most probably indicate an X1 or above class flare. The vertical yellow line shows when the NOAA R3 Radio Blackout alert is raised. Thus, an early initial alert can be raised at the first trigger while the $\Delta EM$ level is monitored continuously. When the $\Delta EM$ increases beyond the minimum threshold an early R3 radio blackout alert can be raised. As can be seen from the left panel, for this flare the first HOPE alert is approximately 10 minutes prior, and the HOPE X-flare alert is 5 minutes prior to the NOAA R3 Radio Blackout alert. The middle panel of Figure \ref{fig:alerts} shows a histogram of the time difference between the standard R3 radio blackout alert time and the time at which the first HOPE flare alert was raised for all category (c) flares analyzed in the previous section. The mean time difference is 6.18 minutes with a standard deviation of 2.76 minutes. The right panel of Figure \ref{fig:alerts} shows a histogram of the time difference between the R3 radio blackout alert and the X-class flare alert raised using the HOPE-based nowcasting method. The mean of this time difference is 2.21 minutes with a standard deviation of 1.59 minutes. This indicates that the HOPE based nowcasting method can provide early alerts, which could be useful for various space weather operations. For example, such an earlier alert could be useful for aircraft operators to be better prepared that they are about to lose High Frequency (HF) radio communications.   

Such a nowcasting technique could also be helpful for flare observation campaigns that aim to observe the impulsive phase of a flare. For example the Extreme-Ultraviolet Variability Experiment (EVE) instrument aboard the SDO mission, runs realtime an algorithm to trigger flare observations for its Multiple EUV Grating Spectrographs B-channel (MEGS-B) \citep{woods_revealing_2025}. However, the current triggering methodology for the MEGS-B flare campaings is based on the magnitude and slope of the SXR timeseries, which misses the flare onset and beginning of the impulsive phase. The EVE SXR measurement is made by its Extreme-ultraviolet SpectroPhotometer (ESP) quadrant-diode channel that has a passband between 1 nm and 7 nm. The ESP 1-7 nm bandpass is sensitive to the hot coronal emissions mainly because of the 2 nm irradiance changes significantly during flares. Previous work (for e.g., \cite{hock_using_2013}), has shown the feasibility of using EVE data as a proxy for GOES XRS channel-B irradiance measurement. Using the HOPE-based nowcasting technique could enable an earlier start time of the flare campaign, which would enable capturing the flare initiation. One way to achieve this is to run the HOPE nowcasting tool on the ground to generate an early alert, which can then be uplinked to the spacecraft to start the flare campaign. It might also be possible for the EVE flight software to use another EVE ESP channel with the ESP SXR channel to estimate the plasma temperature and EM employing a similar flux-ratio technique used for the two GOES-XRS bands. 

Another possible application for this nowcasting technique could be for sounding rockets that aim to capture the impulsive phase of the flare. Such campaigns could start the launch countdown at the first trigger, and decide to launch or not when the $\Delta EM$ value crosses a certain level (which could be tuned for M or X class flares). As shown in the previous histograms, this method could give an advanced alert of a few minutes, which is significant time for a typical sounding rocket to reach observation altitudes (e.g., a NASA Terrier Black-Brant sounding rocket takes about 90 seconds to reach a 100 km altitude). The operational version of this nowcasting algorithm is beyond the scope of this paper, but results shown here indicate a promising early-alert nowcasting methodology. The next section describes a few ideas for future work to improve flare nowcasting.

\section{Future Work}\label{sec:Future_Work}
We suggest two areas for immediate future work, the first in improving the prediction algorithm and the second in developing new instrumentation. A possible improvement of the algorithm could be using a Machine Learning (ML) model, in addition to the physics-based onset temperature and emission measure algorithm developed in this study. The ML model can be trained on the vast GOES dataset spanning decades, to forecast the flare peak magnitude and time. Additional details of the ML model are beyond the scope of this study, but work is underway for the development and training of this model with lessons learned by the HOPE technique. Such a method would incorporate time-histories of the different plasma parameters, rather than using just one data point as done in this study. 

These analyses have been done using full-disk solar irradiance data, but there is the option to apply the HOPE technique to individual pixels of solar EUV or SXR images observed at two different passbands to be able to predict the location of the flare too. Because existing solar EUV imagers do not generally provide a measure of hot plasma temperature above 5 MK, as is needed for the HOPE technique, the second suggestion for an area of future work is the development of new instrumentation. A key limitation in this work is the use of disk-integrated ``sun-as-a-star" measurements for the calculation of the flaring plasma temperature and emission measure. However, because solar flares are localized phenomena, having a SXR imaging spectrometer would be ideal to determine plasma temperature and emission measure for a localized flaring plasma source. An imaging spectrometer would also solve the problem of background subtraction which is a challenge to current analysis, as different flares occur at different background levels. With such an instrument, if the plasma temperature and emission measure is available for each pixel, the flare nowcasting techniques developed in this study could be applied to predict flare location on the solar surface as well. The application of the HOPE based nowcasting algorithm doesn't require high spatial resolution, and thus allows for the possibility of developing a space weather oriented CubeSat/Small-satellite mission  for spatially-resolved solar flare nowcasting.

\section{Data Source}
The DAXSS solar SXR spectral irradiance data products, along with the DAXSS RMF and ARF calibration files, user guide, and data plotting examples in IDL and Python, are available from the MinXSS website at
\url{http://lasp.colorado.edu/home/minxss/}. The GOES XRS data are available from
\url{http://www.ngdc.noaa.gov/stp/satellite/goes/dataaccess.html}. 

\appendix

\section{Parameter Selection for the HOPE-based Nowcasting Algorithm}
\label{sec:appendix}
Section 2.4 described the HOPE-based nowcasting algorithm and the different parameters used in the algorithm. This section describes the analysis performed to find the best set of algorithm parameters. The methodology followed for this analysis was to run the nowcasting algorithm for the same dataset of 137 flares used in this study, iterating through different combination of parameters to find the set that predicts flare alerts most accurately. For each run, the input parameters and the nowcasting algorithm outputs are then stored. Table \ref{tab:hope_param_space} lists the parameter space of all the parameters explored. This results in a total of 135 different model parameter combinations, which were performed and outputs analyzed to find the best model parameters. Three outputs are considered as the main metric for evaluating the performance of a set of parameters, which are total true positive rate (or Recall = True Positives/(True Positives + False Negatives), correlation ($R^2$) between the running-difference emission measure ($\Delta EM$ at local maximum of $\Delta^2 EM$) and the flare peak flux magnitude, and the normalized mean time difference between the first HOPE trigger and the peak flux time (denoted my $T_{A,X}$, $T_{A,M}$, $T_{A,C}$, i.e., mean alert time for X-class, M-class, and C-class flares respectively). 
\begin{deluxetable}{lll}
\tablecaption{Parameter space used for tuning the HOPE nowcasting algorithm \label{tab:hope_param_space}}
\tablehead{
\colhead{Parameter} & \colhead{Symbol} & \colhead{Values}
}
\startdata
Integration Time (seconds)        & $t_{i}$             & 10, 30, 60 \\
Running-difference Time           & $t_{d}$             & $T_{i}$, $3T_{i}$, $5T_{i}$ \\
Emission Measure Threshold        & $\Delta EM$         & 0.001, 0.0025, 0.005, 0.0075, 0.01 \\
Temperature Threshold (MK)        & $\Delta T_{\mathrm{min}}$ & 5, 7.5, 10 \\
\enddata
\end{deluxetable}

\begin{figure*}[!htbp]
\centering
\includegraphics[width = \textwidth]{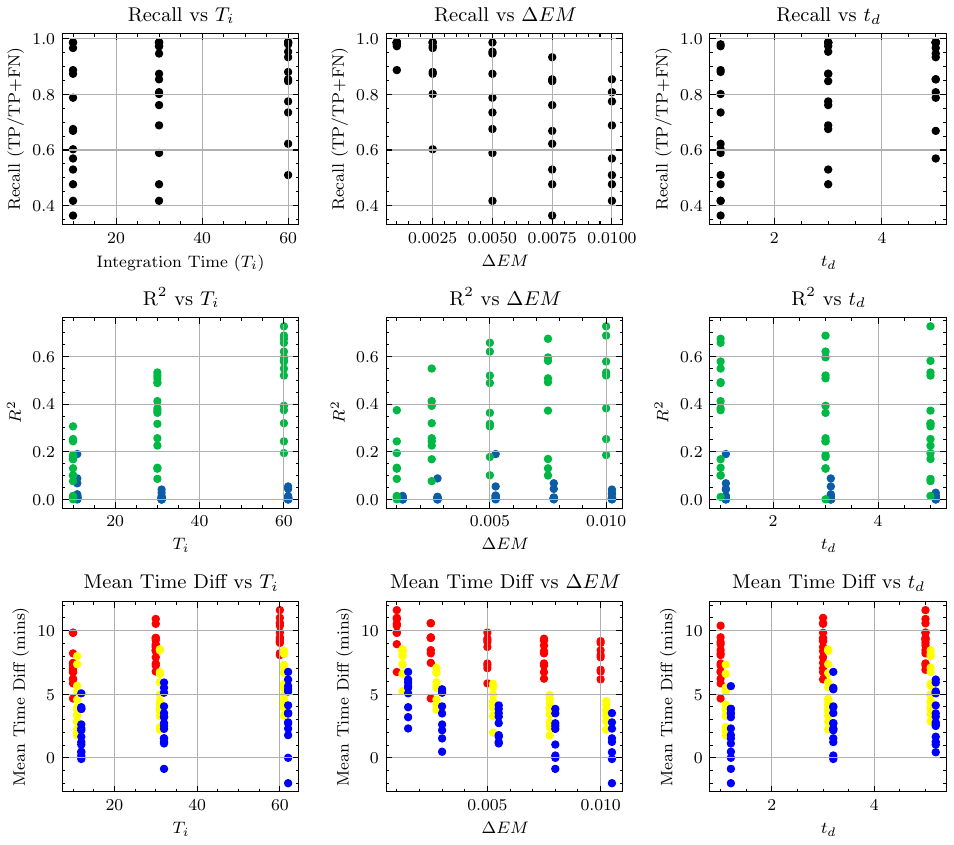}
\caption{The figure shows the variation of the model output metrics for different model parameters. The first row shows the variation of the total recall value (black symbols) with integration time ($t_i$), emission measure threshold ($\Delta EM$), and running-difference time ($t_d$) in the first, second, and third column respectively. The second row shows the variation of the correlation between the value of $\Delta EM$ and the same model parameters. This correlation is computed is shown for two points in the timeseries, the first trigger (blue) and at the local maximum of $\Delta^2 EM$ (green). The blue symbols are offset slightly in the x-axis so that they are visible, even though the blue and green symbols correspond to the same x value. The third row shows the variation of the mean of the time difference (in minutes) between the flare peak and the first trigger. This is shown for C-class flares using blue symbols, M-class flares using yellow symbols, and X-class flares using red symbols. The yellow and blue are symbols are slightly offset in the x-axis so that they are clearly visible, even though they are at the same x-value.}
\label{fig:tuning}
\end{figure*}

Figure \ref{fig:tuning} shows the variation of these output parameters with different model parameters. The first row shows the variation of Recall with integration time, emission measure threshold, and running-difference time in the first, second, and third columns respectively. Similarly, the second row shows the variation of $R^2$ and the third row shows the variation of mean time difference for X (red), M (yellow), and C (blue) class flares. From these plots an integration time of 60 seconds gives good results on all output parameters. A smaller emission measure threshold increases Recall but leads to lower $R^2$. Lastly, the running-difference time shows similar results for the parameter space explored. 
 It was observed that the Temperature Threshold ($\Delta T_{min}$) was not affecting the output metrics significantly so that has been omitted from Figure \ref{fig:tuning}, and the value of $\Delta T_{min} = 5 MK$ was chosen for this study. To select the best performing input parameters, a score was assigned to each performance based on the $R^2$, $\Delta EM$, and normalized mean time difference values. Each output is weighted by a chosen weighting factor, according to the formula below:
\begin{equation}
    S = w_{recall}*Recall + w_{R^2}*R^2 + w_{T_{A,X}}*T_{A,X} + w_{T_{A,M}}*T_{A,M} + w_{T_{A,C}}*T_{A,C}
\end{equation}
where $w$ represent the weighting factors for the difference parameters. For this study, we used $w_{recall} = 0.5$, $w_{R^2} = 0.4$, and $ w_{T_{A,X}}= w_{T_{A,M}}= w_{T_{A,C}}=0.033$. We note that for different nowcasting algorithms, a different scoring methodology can be selected to give more importance to one output rather than another (for example, enhancing time difference of the first alert as compared to the total recall). Table \ref{tab:hope_tuning_results} shows the input parameters and the outputs for all model runs. Since $T_{min}$ was fixed to 5MK, 45 model runs were performed. One of the model runs led to no C class flares being detected so it has been omitted from the table; hence Table \ref{tab:hope_tuning_results} has 44 rows. The score column is color coded to highlight the greater than 75th percentile (green), between 25th and 75th percentile (blue), and less than 25th percentile (red) scores. The row with the highest score (=0.79, highlighted in bold) was used for this study, with the input parameters of $t_i = 60$ seconds, $t_d = 3t_i$, $\Delta EM = 0.005$, and $\Delta T_{min} = 5 MK$ that resulted in a Total Recall of 0.95, an Emission Measure correlation ($R^2$) of 0.62, and a mean time difference for X-class flares of 9.38 minutes.

\begin{deluxetable}{ccccccccccc}[hb!]
\tablecaption{HOPE nowcasting method parameter tuning results for 44 different parameter combinations. \label{tab:hope_tuning_results}}
\tablehead{
\colhead{S.No.} & \colhead{$T_i$} & \colhead{$t_d$} & \colhead{$\Delta EM$} & \colhead{Recall} & \colhead{$R^2$} & \colhead{$T_{A,X}$} & \colhead{$T_{A,M}$} & \colhead{$T_{A,C}$} & \colhead{Score} & \colhead{Ranking}
}
\startdata
1 & 10.00 & 1.00 & 0.0010 & 0.89 & 0.01 & 0.58 & 0.61 & 0.34 & 0.50 & Low \\
2 & 10.00 & 1.00 & 0.0025 & 0.60 & 0.17 & 0.40 & 0.44 & 0.07 & 0.40 & Low\\
3 & 10.00 & 1.00 & 0.0050 & 0.42 & 0.10 & 0.50 & 0.23 & 0.57 & 0.29 & Low\\
4 & 10.00 & 1.00 & 0.0075 & 0.36 & 0.10 & 0.54 & 0.21 & 0.57 & 0.27 & Low\\
5 & 10.00 & 3.00 & 0.0010 & 0.99 & 0.00 & 0.85 & 0.86 & 0.59 &0.57 & Medium\\
6 & 10.00 & 3.00 & 0.0025 & 0.87 & 0.24 & 0.64 & 0.54 & 0.32 &0.58 & Medium \\
7 & 10.00 & 3.00 & 0.0050 & 0.68 & 0.18 & 0.61 & 0.39 & 0.18 & 0.45 & Low\\
8 & 10.00 & 3.00 & 0.0075 & 0.53 & 0.13 & 0.58 & 0.33 & 0.03 & 0.35 & Low\\
9 & 10.00 & 3.00 & 0.0100 & 0.48 & 0.19 & 0.53 & 0.26 & -0.01 & 0.34 & Low\\
10 & 10.00 & 5.00 & 0.0010 & 0.99 & 0.01 & 0.85 & 0.94 & 0.75 &0.58 & Medium\\
11 & 10.00 & 5.00 & 0.0025 & 0.97 & 0.08 & 0.71 & 0.66 & 0.39 &0.57 & Medium\\
12 & 10.00 & 5.00 & 0.0050 & 0.79 & 0.31 & 0.63 & 0.46 & 0.25 &0.56 & Medium \\
13 & 10.00 & 5.00 & 0.0075 & 0.67 & 0.17 & 0.64 & 0.40 & 0.15 & 0.44 & Low\\
14 & 10.00 & 5.00 & 0.0100 & 0.57 & 0.25 & 0.60 & 0.34 & 0.07 & 0.42 & Low\\
15 & 30.00 & 1.00 & 0.0010 & 0.97 & 0.13 & 0.77 & 0.77 & 0.47 &0.61 & Medium \\
16 & 30.00 & 1.00 & 0.0025 & 0.80 & 0.41 & 0.73 & 0.49 & 0.23 &0.61 & Medium\\
17 & 30.00 & 1.00 & 0.0050 & 0.59 & 0.49 & 0.64 & 0.38 & 0.17 &0.53 & Medium\\
18 & 30.00 & 1.00 & 0.0075 & 0.48 & 0.49 & 0.62 & 0.28 & -0.13 & 0.46 & Low\\
19 & 30.00 & 1.00 & 0.0100 & 0.42 & 0.38 & 0.58 & 0.26 & 0.52 & 0.41 & Low\\
20 & 30.00 & 3.00 & 0.0010 & 0.99 & 0.13 & 0.91 & 1.00 & 0.82 &0.63 & Medium\\
21 & 30.00 & 3.00 & 0.0025 & 0.97 & 0.26 & 0.81 & 0.70 & 0.60 &0.66 & Medium\\
22 & 30.00 & 3.00 & 0.0050 & 0.87 & 0.36 & 0.75 & 0.52 & 0.40 &0.64 & Medium\\
23 & 30.00 & 3.00 & 0.0075 & 0.76 & 0.51 & 0.72 & 0.45 & 0.34 &0.63 & Medium\\
24 & 30.00 & 3.00 & 0.0100 & 0.69 & 0.52 & 0.68 & 0.43 & 0.23 &0.60 & Medium\\
25 & 30.00 & 5.00 & 0.0010 & 0.99 & 0.09 & 0.94 & 0.99 & 0.88 &0.62 & Medium\\
26 & 30.00 & 5.00 & 0.0025 & 0.99 & 0.23 & 0.81 & 0.79 & 0.76 &0.66 & Medium\\
27 & 30.00 & 5.00 & 0.0050 & 0.95 & 0.32 & 0.80 & 0.58 & 0.48 &0.66 & Medium\\
28 & 30.00 & 5.00 & 0.0075 & 0.85 & 0.37 & 0.76 & 0.50 & 0.37 &0.63 & Medium\\
29 & 30.00 & 5.00 & 0.0100 & 0.81 & 0.53 & 0.73 & 0.45 & 0.19 &0.66 & Medium\\
30 & 60.00 & 1.00 & 0.0010 & 0.98 & 0.37 & 0.89 & 0.86 & 0.83 & 0.73 & High\\
31 & 60.00 & 1.00 & 0.0025 & 0.88 & 0.55 & 0.82 & 0.65 & 0.52 & 0.73 & High\\
32 & 60.00 & 1.00 & 0.0050 & 0.74 & 0.66 & 0.78 & 0.49 & 0.26 &0.68 & Medium\\
33 & 60.00 & 1.00 & 0.0075 & 0.62 & 0.67 & 0.71 & 0.46 & 0.00 &0.62 & Medium\\
34 & 60.00 & 1.00 & 0.0100 & 0.51 & 0.58 & 0.70 & 0.39 & -0.30 &0.51 & Medium\\
35 & 60.00 & 3.00 & 0.0010 & 0.99 & 0.24 & 0.95 & 0.99 & 1.00 & 0.69 & High\\
36 & 60.00 & 3.00 & 0.0025 & 0.99 & 0.39 & 0.91 & 0.78 & 0.80 & 0.73 & High\\
37 & 60.00 & 3.00 & 0.0050 & 0.95 & 0.62 & 0.81 & 0.65 & 0.52 & 0.79 & *Best*\\
38 & 60.00 & 3.00 & 0.0075 & 0.85 & 0.60 & 0.79 & 0.54 & 0.40 & 0.72 & High\\
39 & 60.00 & 3.00 & 0.0100 & 0.77 & 0.69 & 0.79 & 0.49 & 0.34 & 0.72 & High\\
40 & 60.00 & 5.00 & 0.0010 & 0.99 & 0.19 & 1.00 & 0.96 & 0.91 &0.67 & Medium\\
41 & 60.00 & 5.00 & 0.0025 & 0.99 & 0.32 & 0.91 & 0.83 & 0.77 & 0.70 & High\\
42 & 60.00 & 5.00 & 0.0050 & 0.99 & 0.52 & 0.85 & 0.68 & 0.61 & 0.77 & High\\
43 & 60.00 & 5.00 & 0.0075 & 0.93 & 0.58 & 0.81 & 0.58 & 0.52 & 0.76 & High\\
44 & 60.00 & 5.00 & 0.0100 & 0.85 & 0.73 & 0.78 & 0.52 & 0.41 & 0.77 & High\\
\enddata
\tablecomments{The table shows the model inputs (Integration Time ($T_i$), Running Difference Time ($t_d$), EM Threshold ($\Delta EM$)) and outputs (Total Recall, Combined R$^2$, Normalized Time Differences ($T_{A,X}$, $T_{A,M}$, $T_{A,C}$)) for each parameter combination. The ranking column depicts the percentile scores as follows, 75th percentile (High), between 25th and 75th percentile (Medium), and less than 25th percentile (Low). The row with the highest score is marked as *Best* in the ranking column.}
\end{deluxetable}

\bibliography{references_zotero}{}

\begin{thebibliography}{}
\expandafter\ifx\csname natexlab\endcsname\relax\def\natexlab#1{#1}\fi
\providecommand{\url}[1]{\href{#1}{#1}}
\providecommand{\dodoi}[1]{doi:~\href{http://doi.org/#1}{\nolinkurl{#1}}}
\providecommand{\doeprint}[1]{\href{http://ascl.net/#1}{\nolinkurl{http://ascl.net/#1}}}
\providecommand{\doarXiv}[1]{\href{https://arxiv.org/abs/#1}{\nolinkurl{https://arxiv.org/abs/#1}}}

\bibitem[{K.~A. Arnaud(1996)Arnaud}]{arnaud_xspec_1996}
Arnaud, K.~A. 1996, \bibinfo{title}{{XSPEC}: {The} {First} {Ten} {Years},} 101, 17.
\newblock \url{https://ui.adsabs.harvard.edu/abs/1996ASPC..101...17A}

\bibitem[{A.~F. Battaglia {et~al.}(2023)Battaglia, Hudson, Warmuth, Collier, Jeffrey, Caspi, Dickson, Saqri, Purkhart, Veronig, Harra, \& Krucker}]{battaglia_existence_2023}
Battaglia, A.~F., Hudson, H., Warmuth, A., {et~al.} 2023, \bibinfo{title}{The existence of hot {X}-ray onsets in solar flares,} Astronomy \& Astrophysics, 679, A139, \dodoi{10.1051/0004-6361/202347706}

\bibitem[{J.~C. Chancellor {et~al.}(2014)Chancellor, Scott, \& Sutton}]{chancellor_space_2014}
Chancellor, J.~C., Scott, G. B.~I., \& Sutton, J.~P. 2014, \bibinfo{title}{Space {Radiation}: {The} {Number} {One} {Risk} to {Astronaut} {Health} beyond {Low} {Earth} {Orbit},} Life : Open Access Journal, 4, 491, \dodoi{10.3390/life4030491}

\bibitem[{A. Chandran {et~al.}(2021)Chandran, Fang, Chang, Hari, Woods, Chao, Kohnert, Verma, Boyajian, Duann, Evonosky, Kompella, Tsai-Lin, Kumar, Srivastava, Schwab, Sewell, \& Sarpotdar}]{chandran_inspiresat-1_2021}
Chandran, A., Fang, T.-W., Chang, L., {et~al.} 2021, \bibinfo{title}{The {INSPIRESat}-1: {Mission}, science, and engineering,} Advances in Space Research, 68, 2616, \dodoi{10.1016/j.asr.2021.06.025}

\bibitem[{J.~J. Curto(2020)Curto}]{curto_geomagnetic_2020}
Curto, J.~J. 2020, \bibinfo{title}{Geomagnetic solar flare effects: a review,} Journal of Space Weather and Space Climate, 10, 27, \dodoi{10.1051/swsc/2020027}

\bibitem[{D.~F. da Silva {et~al.}(2023)da Silva, Hui, Simões, Valio, Costa, Hudson, Fletcher, Hayes, \& Hannah}]{dasilva_statistical_2023}
da Silva, D.~F., Hui, L., Simões, P. J.~A., {et~al.} 2023, \bibinfo{title}{Statistical analysis of the onset temperature of solar flares in 2010–2011,} Monthly Notices of the Royal Astronomical Society, 525, 4143, \dodoi{10.1093/mnras/stad2244}

\bibitem[{A.~R. Foster {et~al.}(2012)Foster, Ji, Smith, \& Brickhouse}]{foster_updated_2012}
Foster, A.~R., Ji, L., Smith, R.~K., \& Brickhouse, N.~S. 2012, \bibinfo{title}{{UPDATED} {ATOMIC} {DATA} {AND} {CALCULATIONS} {FOR} {X}-{RAY} {SPECTROSCOPY},} The Astrophysical Journal, 756, 128, \dodoi{10.1088/0004-637X/756/2/128}

\bibitem[{R.~A. Hock {et~al.}(2013)Hock, Woodraska, \& Woods}]{hock_using_2013}
Hock, R.~A., Woodraska, D., \& Woods, T.~N. 2013, \bibinfo{title}{Using {SDO} {EVE} data as a proxy for {GOES} {XRS} {B} 1–8 angstrom,} Space Weather, 11, 262, \dodoi{10.1002/swe.20042}

\bibitem[{H. Hudson(2025)Hudson}]{hudson_anticipating_2025}
Hudson, H. 2025, \bibinfo{title}{Anticipating {Solar} {Flares},} Solar Physics, 300, 2, \dodoi{10.1007/s11207-024-02418-4}

\bibitem[{H.~S. Hudson {et~al.}(2021)Hudson, Simões, Fletcher, Hayes, \& Hannah}]{hudson_hot_2021}
Hudson, H.~S., Simões, P. J.~A., Fletcher, L., Hayes, L.~A., \& Hannah, I.~G. 2021, \bibinfo{title}{Hot {X}-ray onsets of solar flares,} Monthly Notices of the Royal Astronomical Society, 501, 1273, \dodoi{10.1093/mnras/staa3664}

\bibitem[{Y. Kawabata {et~al.}(2018)Kawabata, Iida, Doi, Akiyama, Yashiro, \& Shimizu}]{kawabata_statistical_2018}
Kawabata, Y., Iida, Y., Doi, T., {et~al.} 2018, \bibinfo{title}{Statistical {Relation} between {Solar} {Flares} and {Coronal} {Mass} {Ejections} with {Respect} to {Sigmoidal} {Structures} in {Active} {Regions},} The Astrophysical Journal, 869, 99, \dodoi{10.3847/1538-4357/aaebfc}

\bibitem[{S. Krucker {et~al.}(2020)Krucker, Hurford, Grimm, Kögl, Gröbelbauer, Etesi, Casadei, Csillaghy, Benz, Arnold, Molendini, Orleanski, Schori, Xiao, Kuhar, Hochmuth, Felix, Schramka, Marcin, Kobler, Iseli, Dreier, Wiehl, Kleint, Battaglia, Lastufka, Sathiapal, Lapadula, Bednarzik, Birrer, Stutz, Wild, Marone, Skup, Cichocki, Ber, Rutkowski, Bujwan, Juchnikowski, Winkler, Darmetko, Michalska, Seweryn, Białek, Osica, Sylwester, Kowalinski, Ścisłowski, Siarkowski, Stęślicki, Mrozek, Podgórski, Meuris, Limousin, Gevin, Mer, Brun, Strugarek, Vilmer, Musset, Maksimović, Fárník, Kozáček, Kašparová, Mann, Önel, Warmuth, Rendtel, Anderson, Bauer, Dionies, Paschke, Plüschke, Woche, Schuller, Veronig, Dickson, Gallagher, Maloney, Bloomfield, Piana, Massone, Benvenuto, Massa, Schwartz, Dennis, Beek, Rodríguez-Pacheco, \& Lin}]{krucker_spectrometertelescope_2020}
Krucker, S., Hurford, G.~J., Grimm, O., {et~al.} 2020, \bibinfo{title}{The {Spectrometer}/{Telescope} for {Imaging} {X}-rays ({STIX}),} Astronomy \& Astrophysics, 642, A15, \dodoi{10.1051/0004-6361/201937362}

\bibitem[{R. Lin {et~al.}(2002)Lin, Dennis, Hurford, Smith, Zehnder, Harvey, Curtis, Pankow, Turin, Bester, Csillaghy, Lewis, Madden, van Beek, Appleby, Raudorf, McTiernan, Ramaty, Schmahl, Schwartz, Krucker, Abiad, Quinn, Berg, Hashii, Sterling, Jackson, Pratt, Campbell, Malone, Landis, Barrington-Leigh, Slassi-Sennou, Cork, Clark, Amato, Orwig, Boyle, Banks, Shirey, Tolbert, Zarro, Snow, Thomsen, Henneck, Mchedlishvili, Ming, Fivian, Jordan, Wanner, Crubb, Preble, Matranga, Benz, Hudson, Canfield, Holman, Crannell, Kosugi, Emslie, Vilmer, Brown, Johns-Krull, Aschwanden, Metcalf, \& Conway}]{lin_reuven_2002}
Lin, R., Dennis, B., Hurford, G., {et~al.} 2002, \bibinfo{title}{The {Reuven} {Ramaty} {High}-{Energy} {Solar} {Spectroscopic} {Imager} ({RHESSI}),} Solar Physics, 210, 3, \dodoi{10.1023/A:1022428818870}

\bibitem[{S. Nagasawa {et~al.}(2022)Nagasawa, Kawate, Narukage, Takahashi, Caspi, \& Woods}]{nagasawa_study_2022}
Nagasawa, S., Kawate, T., Narukage, N., {et~al.} 2022, \bibinfo{title}{Study of {Time} {Evolution} of {Thermal} and {Nonthermal} {Emission} from an {M}-class {Solar} {Flare},} The Astrophysical Journal, 933, 173, \dodoi{10.3847/1538-4357/ac7532}

\bibitem[{A. Telikicherla {et~al.}(2024)Telikicherla, Woods, \& Schwab}]{telikicherla_investigating_2024}
Telikicherla, A., Woods, T.~N., \& Schwab, B.~D. 2024, \bibinfo{title}{Investigating the {Soft} {X}-{Ray} {Spectra} of {Solar} {Flare} {Onsets},} The Astrophysical Journal, 966, 198, \dodoi{10.3847/1538-4357/ad37f6}

\bibitem[{S.~M. White {et~al.}(2005)White, Thomas, \& Schwartz}]{white_updated_2005}
White, S.~M., Thomas, R.~J., \& Schwartz, R.~A. 2005, \bibinfo{title}{Updated {Expressions} for {Determining} {Temperatures} and {Emission} {Measures} from {Goes} {Soft} {X}-{Ray} {Measurements},} Solar Physics, 227, 231, \dodoi{10.1007/s11207-005-2445-z}

\bibitem[{T.~N. Woods {et~al.}(2024)Woods, Eden, Eparvier, Jones, Woodraska, Chamberlin, \& Machol}]{woods_goes-r_2024}
Woods, T.~N., Eden, T., Eparvier, F.~G., {et~al.} 2024, \bibinfo{title}{{GOES}-{R} {Series} {X}-{Ray} {Sensor} ({XRS}): 1. {Design} and {Pre}-{Flight} {Calibration},} Journal of Geophysical Research: Space Physics, 129, e2024JA032925, \dodoi{10.1029/2024JA032925}

\bibitem[{T.~N. Woods {et~al.}(2023)Woods, Schwab, Sewell, Kandala, Mason, Caspi, Eden, Chandran, Chamberlin, Jones, Kohnert, Moore, Solomon, \& Warren}]{woods_first_2023}
Woods, T.~N., Schwab, B., Sewell, R., {et~al.} 2023, \bibinfo{title}{First {Results} for {Solar} {Soft} {X}-ray {Irradiance} {Measurements} from the {Third} {Generation} {Miniature} {X}-{Ray} {Solar} {Spectrometer},} The Astrophysical Journal, 956, 94, \dodoi{10.3847/1538-4357/acef13}

\bibitem[{T.~N. Woods {et~al.}(2025)Woods, Chamberlin, Jones, Mason, Qian, Warren, Woodraska, Borelli, Eparvier, \& Gonzalez}]{woods_revealing_2025}
Woods, T.~N., Chamberlin, P.~C., Jones, A., {et~al.} 2025, Revealing {Flare} {Energetics} and {Dynamics} with {SDO} {EVE} {Solar} {Extreme} {Ultraviolet} {Spectral} {Irradiance} {Observations}, arXiv, \dodoi{10.48550/arXiv.2507.19681}

\end{thebibliography}
\bibliographystyle{aasjournalv7}



\end{document}